# On the Proper Use of Statistical Analyses; a Comment on "Evaluation of Colorado Learning Attitudes about Science Survey" by Douglas et al.

Carl E. Wieman

Department of Physics and Graduate School of Education, Stanford University, Stanford, CA 94305

Wendy K. Adams

Science Education Programs and Department of Physics and Astronomy, University of Northern Colorado, Greeley, CO 80639

Abstract

The paper "Evaluation of Colorado Learning Attitudes about Science Survey" [1] proposes a new, much shorter, version of the CLASS based on standard factor analysis. In this comment we explain why we believe the analysis that is used is inappropriate, and the proposed modified CLASS will be measuring something quite different, and less useful, than the original. The CLASS was based on extensive interviews with students and is intended to be a formative measurement of instruction that is probing a much more complex construct and with different goals than what is handled with classic psychometrics. We are writing this comment to reiterate the value of combining techniques of cognitive science with statistical analyses as described in detail in Adams & Wieman, 2011 [2] when developing a test of expert-like thinking for use in formative assessment. This type of approach is also called for by the National Research Council in a recent report [3].

PACS numbers: 01.40Fk, 01.40G-, 01.40.Di, 01.50Kw

In recent years there has been a growing effort to develop assessment tools that target students' development of expert-like thinking, both conceptually and attitudinally. The CLASS [4] is an example of one that measures students' perceptions of physics as a discipline. Other assessment tools probe whether students understand and apply particular concepts in the manner of a scientist in the discipline. Such assessment tools are intended to measure student learning and attitudinal development in courses to provide Formative ASsessment of Instruction (FASI). The methodology involved in developing and validating such assessment tools has been presented in prior work [3, 4].

We will briefly describe here how this approach was used to develop the original widely used, eight category, 42 statement version of the CLASS, and how it differs from the approach of Douglas et al. In addition we will highlight data provided by the CLASS that demonstrates the value of our CLASS category structure. Finally, we will discuss a number of additional methodological problems in the Douglas et al. analysis, although our analysis of this paper was limited by the absence of important information.

**Development and Validation of the CLASS**

The CLASS is a relatively short assessment, even with 42 statements, that takes approximately eight minutes to complete and is typically administered online or in-class. The fundamental evidence of validity for the CLASS was collected through a combination of extensive student interviews and rigorous, albeit at times groundbreaking, statistical techniques. Our work was based upon the extensive previous work on such student perceptions by Hammer, Bransford, Redish and others [5,6,7,8] and the MPEX [9], which was the first survey of this type for undergraduate physics. Interviews with students and physics faculty about physics consistently show that there are certain aspects of physics and learning physics that students bring up and care about and have varying views on, but on which physicists have very consistent views. This difference between many students and professional physicists remains the fundamental criteria for inclusion of a particular perception on the CLASS. These perceptions of student thinking define the focus of the CLASS, but as people's views on as broad a topic as physics and learning physics are complex and multi-faceted, these perceptions do not necessarily correlate well with each other.

We emphasize that what the CLASS measures was defined by these statements that reflected issues of common concern to students and faculty as expressed in numerous interviews, not by any well defined theoretical construct that we set out to measure. As such, it is a fundamentally different entity from the single well-defined construct that is the target of most high stakes formal assessment. These latter type of assessments are the basis for the "assessment fundamentals" section and the subsequent analysis of the Douglas paper. We believe that it is incorrect to expect students' perceptions about the various aspects of physics to be so simply characterized; a belief that is bolstered by our interview data, as discussed below. In addition, the CLASS is intended to serve a fundamentally different purpose than the tests that are the basis of the psychometrics discussion in the Douglas paper. Those tests are intended to be summative measures of student learning that provide maximum discrimination between students. The goal of the CLASS is to provide formative assessment of instruction, to be used by teachers and researchers. This introduces a set of design issues and constraints, such as the need to minimize the class time needed to administer the instrument, that go beyond the usual psychometric considerations.

In the original CLASS paper [4] we had a substantial (admittedly complex and somewhat arcane) discussion about how to create categories; categories which are primarily intended for the benefit of instructors using the CLASS. Our method weighed factor analysis and other statistical analyses along with the criteria of "useful for instructors" and "reflecting the actual way students' brains function as indicated by interview data." In that CLASS paper we were focused on arguing for requiring categories to have some degree of statistical validity to better capture student thinking, rather than being completely based on views of what researchers/instructors might expect, which was the case for the MPEX and other precursors to the CLASS. The method of reduced-basis factor analysis allowed us to do this with a multi-dimensional instrument. We used a direct oblimin rotation which allows for oblique factors. With student perceptions, as in any psychological test, there are factors that are related but still separate; oblique rotation is designed to handle this sort of data [10].  We have several complex variables, statements that load onto more than one factor, but all the statements on each category load evenly and strongly, thereby defining the category.

As has been noted by experts on factor analysis, psychological tests do not fit the idea of independent factors. "…note that when items or facets have complex patterns of loadings, aiming for mathematically based simple structure solutions may not be appropriate. Researchers may need to plot factors and to develop hand rotations (Comrey, 1978; Comrey & Lee, 1992; Gorsuch,1983)[11,12,13]" ([10], p. 292) Thus, our theoretical foundation is not original.   We did create the "robustness rating" which is a single number derived from the multiple criteria recommended for analyzing the strength of a category.

Finally, we also made sure that our resulting categories were consistent with correlations or distinctions that we found in our interview data. Interviews also remained our ultimate test of the validity of the individual CLASS statements.  All 42 statements were consistently interpreted by students and describe their perceptions of physics as a discipline and how best to learn physics.

Our approach takes a more nuanced view of student thinking that relies on more than simple psychometric analysis, in accordance with what is called for in the 'Standards for educational and psychological testing' [14] and the National Research Council's study of and recommendations on educational assessment [3]. Our development methodology follows the guidelines provided in those publications. These two publications are much more thorough, and give far more weight to achieving meaningful goals with the intended assessment, than what can be found in the references cited by Douglas et al.

**Development of measures of expert-like thinking**

In a paper following the development of the CLASS and several other FASIs, including concept inventories, we [2] provided a longer and more detailed discussion about the design criteria and the relevance of various standard psychometric criteria in such contexts. The basic theoretical idea behind all these FASI instruments is that there are certain ways of thinking that experts within a particular subject share, and to be useful a test must accurately capture that expert thinking. The emphasis on the need to ensure the test is correctly probing the degree to which the students are thinking like experts is an increasing focus of educational assessment. Quoting from "Knowing What Students Know. The science and design of educational assessment" ([3] p. 4), '*Studies of expert-novice differences in subject domains illuminate critical features of proficiency that should be the targets for assessment.*' As such, the instrument will often be probing many different facets of thinking and learning, rather than a single construct. However, such a test should also address usefulness to instructors, including defining particular aspects to which instruction can be targeted, as well as the constraints of limited classroom time and student attention for administration. These factors are critically important for a good FASI but are not included in simple psychometric criteria.

One particularly important part of both the development and validation of the CLASS (and other FASIs) is the use of student interviews. There is a large body of literature on the use of student interviews for the purpose of understanding student thinking [15, 16]. Student interviews are rarely used when developing formal educational tests, even though the value of this kind of information is stressed in the 2001 NRC report. '*The methods used in cognitive science to design tasks, observe and analyze cognition, and draw inferences about what a person knows are applicable to many of the challenges of designing effective educational assessments*' ([3], p.5). The Standards [14] have an extensive discussion on the collection of evidence of validity, but only a small fraction of that discussion refers to psychometric tests.

The NRC calls for having three elements in all assessments: *cognition, observation, and interpretation* [3]. The expert novice differences of value to teachers define the cognition that is being probed by the FASI (including CLASS); the questions themselves are the tasks that elicit this thinking (observations); and the validation process - interviews demonstrating that the FASI measures what is intended - determines the interpretation by delineating what inferences about student thinking can be drawn from the results of the FASI.

Also, with any FASI the results from the students as a group are more important than ranking individual students, and they must necessarily be a low stakes assessments to preserve their validity and value. This combination of factors relaxes a number of constraints on test design [3], and it makes the test design and suitable statistical tests of the instrument rather different ([14], p. 13) than what is considered in simple psychometric analysis.

Psychometricians use statistical tests such as item analysis, Item Response Theory, or Factor Analysis to analyse the validity of a test. However, as noted above, the standard acceptable ranges of values for these statistical tests were determined for tests with different goals and design criteria, and are not appropriate for FASIs. Because the CLASS and other formative assessments used in education research serve a much different purpose and have different constraints, different statistical criteria should be used.

As an example, the shorter, 15 statement version of the CLASS in [1] is based on the requirement that student perceptions are a single construct. Douglas et al.'s first step before performing factor analysis was to remove ten statements based only on their correlation coefficients with other CLASS statements. Because the responses to various CLASS statements do not show the relationships called for by a single construct instrument, Douglas et al. conclude they cannot be part of "general attitude about physics" ([1],p. 8), and so should not be on the CLASS. Dropping such statements because they do not fit a single construct is akin to assuming that all cows are spherical and thereby dropping all bovine measures that are not spherically symmetric. It makes for a tidier analysis, but tells you much less about the characteristics of cows, and correspondingly, less about how and why students actually feel the way they do. The way real people feel about the topic of physics cannot be so nicely lumped into a single clean construct. Those "offending" CLASS statements reflect the degree to which a student is more or less like a physicist in their feeling about physics, and those feelings can have a big impact on whether the student is happy or unhappy about a physics course and whether or not they will choose to pursue physics as a career. This is true even if those feelings are not well correlated with other feelings they may have about physics that are probed with the CLASS.

**Analysis supporting the value of the original CLASS categories**

As an example of the limitations of the Douglas et al. approach, consider their proposed category Personal Application and Relation to the World which combines the original CLASS Personal Interest and Real World Connection categories. In our original factor analysis we also found that these two categories showed up as a single large category. However, in student interviews there was a clear difference between a student recognizing that physics applies to the real world and the student caring personally about physics in their world.

As we refined our factor analysis technique, the method evolved into what we labeled as a "reduced-basis factor analysis" in the original CLASS paper. We carried out factor analysis on subsets of statements to reduce distortion of the factors by other statements with low commonalities. This does not imply a lack of value for any such statements, only that there are some factors which overwhelm others, and so analyzing subsets can reveal nuances that are lost when including the full set of statements in a factor analysis. What we discovered is that when all the statements that made up the large combined real world/personal interest category were analyzed without the rest of the CLASS statements, this one large category indeed did split into two separate categories. These two categories nicely matched what had been seen in student interviews.

This distinction between the composite and split categories is important. When analyzing who becomes a physics major, the strongest correlation is with the Personal Interest category, rather than the complete survey score or any other category. Table VIII [4] in the original CLASS paper shows another place where the distinction is important. Four different types of courses were analyzed at two different universities. The difference between the Overall scores for males and females range from 2 to -13%; however, in every case the difference between male and female views in the Personal Interest category was substantially greater, ranging from -11 – -34%. In each case, there is a three to four fold increase in sensitivity to gender differences if one looks at the Personal Interest category alone, rather than the overall score or other categories.

We have also seen a clear and important difference in the results of the Personal Interest and the Real World Connection categories in the pre- and post-semester shifts for different courses. The results in CLASS Table I and Table V from the original paper show that these categories shift differently based on instruction (Table I.) This type of sensitivity is valuable when instructors are trying to understand what they have done well and where they can still improve when addressing student perceptions of physics.

TABLE I. Variation in shifts in different courses

|  | Shift Calc Based at LSRU | Shift Calc Based at MMSU |
|---|---|---|
| Overall | -7 (1)% | -9 (1)% |
| Personal Interest | -11 (1)% | -5 (5)% |
| Real World Connection | -7 (1)% | -23 (7)% |

Shift in percent favorable responses pre to post – semester. Standard error in parentheses.

This illustrates one example of what would be lost using the Douglas et al. version of the CLASS and categories. There are many other similar useful distinctions in our CLASS category divisions which we will not present here, in the interest of space.

It is also important to recognize what would be lost by dropping CLASS statements that do not fall into any category. Particular aspects of the student perceptions were reflected in only one or two CLASS statements. These were found sufficient to capture the aspect in question while avoiding redundancy, particularly as we deliberately removed statements that showed a high correlation with another statement to optimize the use of student time and attention. These "orphan" statements accurately capture important and oft-expressed views, and hence we believe they are important to keep, although these views are not well correlated with responses to other statements, and so do not fall into any category.

One minor point which has caused confusion with the CLASS is the inclusion of statements that are not scored. These mainly concern students' views about studying physics which some instructors indicated were important to them to know, but there is no consistent expert response on these statements, and so we do not score them on the novice-expert scale used with the other CLASS statements.

**Problems with the Douglas et al. paper**

We have a number of additional concerns with the methodology in the Douglas et al. analysis, but our discussion is limited by problems with the material and missing data in the paper itself.[1] Unfortunately much of the data presented in the Tables and the figure are not defined in the paper.

Figure 1, for example, presents the proposed factor structure for the final 15-item version of the CLASS but it does not include many of the statistics that were used by the authors to argue for a smaller set of statements. Additionally, on the right side of Figure 1 each statement is repeated in an oval and in some cases with bi directional arrows and covariances, there are also p-values included with the factor loadings and covariances. There is no description either in the body of the text or the figure caption explaining the significance of these items. Table I is labeled as "Exploratory Factor Analysis" and "Confirmatory Factor Analysis"; however, the data in this table are a summary of various statistics including mean student response, standard deviation, and item total correlations for each statement, apparently for the two data sets used in the factor analysis, but are not factor analysis results. Table II uses S for Factor Structure and P for Pattern Structure. In the text and Table II title "factor structure" is used to mean something different from the coefficients that are found in the table. These numbers are more likely "structure matrix" coefficients. However, nowhere in the text are these defined or referred to when describing how the analysis was performed. The text does refer to factor loadings but does not tell readers which, if either, of these two columns "P" or "S" are what is referred to as factor loadings. Table II also includes a column of data labeled as "$h^2$" but does not define what it is nor its significance to their analysis. The undefined coefficients and the $h^2$

---

[1] Note added. After this comment was submitted, several changes were made in the online published version including the submission of Figure 1 and correction of Table IV showing the final 15 statement version rather than the interim 17 statement version.

index makes interpretation difficult when there is no explanation given as to how or if this additional data is used.

The three new proposed categories/factors in the Douglas et al. paper are inconsistently named, so it is difficult to be certain as to what is being referred to at different places in the paper. See table II for a listing of the names found at different places in the paper for what we are guessing refer to the same categories, but cannot be sure do not refer to different but undefined entities.

Table II: Inconsistent Category Naming

| Factor 1: | Name | Location |
|---|---|---|
| | Personal Application | Abstract, Figure 1 |
| | Personal Interest and Relation to the Real World | Page 5 |
| | Personal Application and Relation to the World | Page 7 |
| | Personal Application and Relation to Real World | Table II & IV |
| **Factor 2:** | | |
| | Problem Solving | Abstract, Figure 1 |
| | Approaches to problem solving | Abstract |
| | Problem Solving and Learning | Page 5, page 7, Table II |
| | Problem Solving/Learning | Table IV |
| **Factor 3:** | | |
| | Personal effort | Abstract, Figure 1 |
| | Effort and Sense Making | Page 5, Table II |
| | Effort/Sense Making | Table IV |
| | Personal Effort and Sense Making | Page 7 |

Many of the references in the Douglas paper are incorrect or inappropriate. Most of these errors are confusing but minor, except for the sentence on page 4 that states the CLASS is based on Fishbein's Theory of attitudes and gives a reference to our original CLASS paper in that sentence. The CLASS is not based on the Fishbein theory, and the CLASS development and validation paper does not reference this work.

Beyond the inconsistent information in the paper, there are some methodological issues that are apparent. First, Douglas et al. also do not offer student interview results to support the validity of this shorter version or the categories that they have proposed. We believe these to be essential tests of validity; a belief that is reiterated in the authoritative references we have listed. Students often can respond to the statements differently if they are presented in a different order so our previous tests of validity of the individual statements may no longer apply.

Second, there are no scored results presented from this shorter version. Without such data it is unjustified to claim that students will show similar responses on the shorter CLASS as they do on that same subset of questions on the full CLASS. Similarly, Douglas et al. have not done a factor analysis on data collected with their shorter version to verify that indeed these factors look the same when the survey is administered in this proposed modified format as they do when analyzing a subset of the full set of responses to the CLASS.

Third, their analysis is based on a demographically narrow set of students that is 77% male, 85% non-minority students. The sample also had a somewhat low response rate of 50%, which opens up the possibility of there being a self- selected bias in this population.

Fourth, the discussion of Table II in the Results section says there are cross loadings for five statements. However, the data shown indicates cross loading for three additional statements, 24, 30 and 40. Some of these are larger cross loadings than found with the statements they list. Cross loadings are not presented for the proposed final 15-item version except for statement 25.

Fifth, the Results section describes how the communality was used to remove any of the CLASS statements with a value lower than 0.3. However, Table II reveals four statements in the 17-item version with communalities lower than 0.3 (Typically $h^2$ is the *communality* of each statement.) Communalities are not presented for the proposed final 15-item version.

Finally, there are inconsistencies between the theoretical framework presented and the actual approach used in the Douglas et al. choices. The authors manually add modification indices to allow them to make personal judgments such as "…allow retention of the statement *I enjoy solving physics problems* which is conceptually an important aspect of a students' attitude about physics and learning physics." Such judgments on their personal "conceptual" grounds are inconsistent with their theoretical framework presented earlier. If that is to be the criteria used, one should keep all the other CLASS statements they previously dropped, as there is a large body of existing interview data indicating that those statements are conceptually important to students and faculty.

**Ongoing validation.**

Douglas et al., do mention how validity is not a one-time stamp and that evidence must be collected for different populations. We certainly agree with that sentiment, which is consistent with the Standards for Educational Testing as well as our FASI paper. It is essential that the students interviewed for both the development and validation of the test represent the breadth of the population for which the test is to be used ([14], Standard 1.4). An exemplary case of this is found in the work of Sawtelle, Brewe and Kramer [17]. The original CLASS was validated with a population of only 20% non-Caucasians. Researchers at Florida International University (FIU) recognized the need for validation with their different student population so interviewed students on the CLASS statements to validate the use of the instrument with Hispanic students, which compose nearly 60% of the FIU student body. They conclude "We find that in our predominately Hispanic population, 94% of the students' interview responses indicate that the students interpret the CLASS items correctly, and thus the CLASS is a valid instrument. We also identify one potentially problematic item in the instrument which one third of the students interviewed consistently misinterpreted" ([17] p. 1). Anyone considering using the CLASS with diverse populations should use the findings of that paper.

**Conclusion**

As physics and other discipline-based education researchers develop new assessment tools, it is important that they understand the value of both statistical techniques and methodology from cognitive science as recommended by the National Research Council and the Standards for Psychological testing [3,14]. Psychometrics and the oft quoted statistical measures of reliability and validity are a part of these development procedures but must be used appropriately. These statistical measures make sense only within a limited context and where certain fundamental assumptions hold. Far too often those underlying assumptions are forgotten or unclear, and criteria get applied where they are not just inappropriate but can be detrimental to the quality of the resulting instrument.

The CLASS was developed with careful consideration of the value and limitations of the statistical techniques used and was strongly informed by student interviews. The resulting instrument has demonstrated high fidelity in understanding student perceptions of physics and the effects of instruction on those perceptions. Using only a third of the original questions and a modified set of categories, would substantially reduce its utility and make comparisons with results of the original CLASS meaningless.